\documentclass[aps,prb,preprint,superscriptaddress,showpacs,endfloats*]{revtex4}
\usepackage{graphicx}
\usepackage{bm}
\usepackage{amsmath}
\newcommand{\AM}{\text{\AA}\ensuremath{^{-1}}}

\begin{document}

\title{Dynamic spin susceptibility of paramagnetic spinel  LiV$_2$O$_4$}

\author{V. Yushankhai}

\affiliation{Max-Planck-Institut f\"ur Physik Komplexer Systeme, D-01187
Dresden, Germany}

\affiliation{Joint Institute for Nuclear Research, 141980 Dubna, Russia}

\author{A. Yaresko}

\affiliation{Max-Planck-Institut f\"ur Physik Komplexer Systeme, D-01187
Dresden, Germany}

\author{P. Fulde}

\affiliation{Max-Planck-Institut f\"ur Physik Komplexer Systeme, D-01187
Dresden, Germany}

\author{P. Thalmeier} 

\affiliation{Max-Planck-Institut f\"ur Chemische Physik fester Stoffe,
D-01187 Dresden, Germany}

\date{\today}

\begin{abstract}
In an attempt to explain inelastic neutron scattering data for LiV$_2$O$_4$
the dynamical spin susceptibility $\chi({\bf Q},\omega)$ at zero
temperature is calculated. Starting point is a weak coupling approach based
on the LDA bandstructure for that material. It is supplemented by a RPA
treatment of local on-site interactions and contains an adjustable
parameter. Due to the geometrically frustrated lattice structure the
magnetic response is strongly enhanced in the vicinity of a nearly
spherical surface in ${\bf Q}$-space. We compare these results with recent
low-temperature neutron scattering data. The measured spin relaxation rate
$\Gamma$ is used to estimate the spin fluctuation contribution to the
specific heat.
\end{abstract}


\maketitle

\section{Introduction}

The metallic spinel LiV$_2$O$_4$ has been identified as the first 3$d$
system with heavy quasiparticles.\cite{Kondo97,Johnston00,Kondo99} The
low temperature specific heat coefficient $\gamma = C/T \simeq$0.4 J
mol$^{-1}$K$^{-2}$ is considerably enhanced as compared with that of simple
metals and the same holds true for the spin susceptibility $\chi_S$. For
high temperatures, $T>$50 K, the latter shows a Curie-Weiss behavior,
$\chi_S{(T)} = \chi_0 + C/(T+\theta)$, where the Curie constant $C$ = 0.47
cm$^3$K/(mol V) and the Weiss temperature $\theta = 63$ K$>0$ implies an
antiferromagnetic interaction between the V ions.  No magnetic ordering has
been observed down to 0.02 K.\cite{Kondo97}
Concerning the entropy $S(T)$ whose temperature dependence is determined
from the specific heat, one finds that $S(T = 60$ K$)-S(0)\simeq$10
J/(mol$\cdot$K) which is close to 2$R\ln$2 where $R$ denotes the gas
constant. Thus at 60 K there is nearly one excitation per V ion. This
suggests strongly that the low-lying excitations which lead to the heavy
quasiparticle behavior result from spin degrees of freedom. Needless to say
that the original discovery of the heavy quasiparticles in LiV$_2$O$_4$ has
initiated many experimental studies to follow including transport, NMR,
\cite{Fujiwara98,Fujiwara04} $\mu$SR,\cite{Koda05} PES,\cite{Shim06} and
inelastic neutron scattering work.\cite{Krimmel99,Lee01,Murani04}  A
special feature of the spinels and hence of LiV$_2$O$_4$ is that the V ions
occupy the sites of a pyrochlore lattice. The latter consists of corner
sharing tetrahedra.  For a microscopic understanding of the observed heavy
quasiparticle behavior with an average number $n_d=1.5$ of $d$-electrons
per V ion, two different approaches starting from either a strong or a weak
coupling limit are possible. They have been reviewed in Ref.\
[\onlinecite{Fulde03}] and we want to discuss them briefly now.

Density functional calculations based on the LDA demonstrate that electrons
near the Fermi energy E$_F$ are of $t_{2g}$
character.\cite{Matsuno99,Singh99,Eyert99} The calculated density of states
within LDA is too small by a factor of 25 in order to explain the large
$\gamma$ coefficient. Therefore in Ref.\ [\onlinecite{Fulde01}] the limit
of strong correlations was taken as a starting point. In that limit, charge
fluctuations between V ions are strongly suppressed due to both the on-site
and nearest neighbor inter-site electron Coulomb
repulsions.\cite{Yushankhai05} Thus, in the ground state slowly varying
charge configurations include V sites that are either singly (spin-1/2) or
doubly (spin-1) occupied. In the latter case S = 1 is due to the Hund's
rule.  The ground state consists of configurations in which the spin-1/2
and spin-1 sites form two subsets of chains (rings) and the spin chains are
effectively decoupled because of geometrical frustration in the pyrochlore
lattice. Within this spin chain model, the nearest neighbor ($nn$) exchange
couplings $J_{nn}$ are assumed only. The linear specific heat coefficient
$\gamma=2k_{\text{B}}R/3J_{nn}$, where $k_{\text{B}}$ is the Boltzmann
constant, of the spin-1/2 chains is found to be large.  Only a factor of
two is missing as compared with experiment.  Here $J_{nn}\simeq$3 meV is
the $nn$ spin-1/2 chain coupling, as obtained from LDA+U calculations. The
spin-1 chains with a gapped spectrum do not contribute to $\gamma$
coefficient. Our recent analysis has shown, however, that in the model
formulation the next-neares ($nnn$) coupling $J_{nnn}$ cannot be excluded,
which may lead to a considerable renormalization of the above estimate of
$\gamma$. Moreover, if $J_{nnn}/J_{nn}>0.24$, the spin-1/2 chain system may
even enter the regime of dimerization. Of course, effects of the kinetic
energy which were omitted so far should also be included in an extended
spin chain model.

Because of these growing complications, it is of interest to investigate
what results for the spin correlations are, when one starts from the limit
of weak correlations instead. In this limit, the $d$-electrons are in broad
LDA energy bands in contrast to nearly localized electronic states in the
opposite limit.  Since the effects of electron Coulomb repulsions are
neglected (except for those already contained in the LDA), the
geometrically frustrated lattice structure plays a role only as it affects
the form of the LDA energy bands. As shown below, this results however in a
characteristic form of the ${\bf Q}$-dependence of the unenhanced dynamic
spin susceptibility $\chi^{(0)}({\bf Q},\omega)$. So the starting points
of both approaches are very different. The hope is that both limits can
eventually brought to convergence in a region which is between the two
limiting situations.

The present work was first motivated by results of quasielastic neutron
scattering measurements\cite{Krimmel99,Lee01,Murani04} carried out on
polycrystalline samples of LiV$_2$O$_4$. At low temperatures strong
quasielastic neutron scattering was found in a range of wavevectors,
0.4$\AM \lesssim Q\lesssim 0.8\AM$. The quasielastic linewidth $\Gamma(Q)$
is of a few meV, indicating a slow spin dynamics in the above $Q$
region. Because of the relation between the measured dynamic structure
factor and the imaginary part of the dynamic spin susceptibility $S({\bf
Q},\omega;T)=\left(1-e^{-\hbar\omega/k_{B}T}\right) \chi^{\prime
\prime}({\bf Q},\omega;T)$, calculation of $\chi({\bf Q},\omega;T)$ is of
fundamental importance to interpret inelastic neutron scattering data.

Our aim is to include a realistic electronic band structure for
LiV$_2$O$_4$ into RPA-like calculations of the dynamic spin
susceptibility. The latter is enhanced due to exchange-correlation effects
of electrons on the vanadium 3$d$-orbitals. The calculations are performed
for $T$=0. We restrict ourselves to the study of $\chi({\bf Q},\omega)$
along high-symmetry directions in the reciprocal space and focus on the
analysis of its low frequency behavior. As shown below, even such a
restricted consideration allows us to gain insight into the low-temperature
spin dynamics of LiV$_2$O$_4$ and provides a basis for a comparison of
computational results with experimental data. Our main findings are the
following (i) the dominant maxima of the calculated unenhanced spin
susceptibility $\chi^{(0)}({\bf Q},\omega)$, i.e., the one from LDA occur
in the same region of ${\bf Q}$-space where the largest intensity of the
inelastic neutron scattering is found; (ii) a moderate value of the
electronic local exchange-correlation coupling ${\mathcal{K}}$ is
sufficient within RPA to bring the system close to a magnetic instability
and thus to a strong slowing down of spin fluctuations for wave vectors
${\bf Q}$ determined by (i).

Because we assume local, Hubbard-like electron interactions, the
exchange-correlation coupling ${\mathcal{K}}$ is ${\bf Q}$-independent and
therefore treated here as an orbital-independent adjustable
parameter. Along the way, we will discuss possible extensions of the theory
in some detail, but not pursue them further.

The random-phase approximation was applied to the one-band Hubbard model by
Izuyama, Kim, and Kubo in their seminal work\cite{Izuyama63} and developed
further by Doniach.\cite{Doniach67} The multiband generalization of the RPA
to the dynamic spin susceptibility adopted here is in a close relation with
earlier works by Cook\cite{Cook73} and Callaway and
co-workers.\cite{Callaway75,Callaway83} We mention also papers by Stenzel
and Winter \cite{Stenzel85,Stenzel86} along that line. The connection with
these works will be pointed out at various places in the discussion below.

We address also the problem of spin fluctuation contribution $\gamma_{sf}$
to the low temperature specific heat coefficient $\gamma$. As
known,\cite{Lonzarich86,Lonzarich89,Konno87} slow spin-fluctuation dynamics
can be described approximately by a system of overdamped oscillators.  In
general, these oscillators are characterized by ${\bf Q}$-dependent spin
relaxation rates $\Gamma_{\bf Q}$. Based on the available inelastic neutron
scattering data \cite{Krimmel99,Lee01,Murani04} and our RPA calculation of
$\chi({\bf Q},\omega)$ in LiV$_2$O$_4$, we suggest below a model describing
a particular distribution of $\Gamma_{\bf Q}$ in ${\bf Q}$-space.  Then we
show that a diversity of experimentally determined values of $\Gamma_{\bf
Q}$ leads to a theoretical estimate for $\gamma_{sf}$ that falls also into
a broad range with the largest value being close to $\gamma$ observed in
experiment.

A similar RPA approach to a fluctuation mechanism in LiV$_2$O$_4$ based on
a tight binding model for the V-3d bands has been proposed in
Ref.\ [\onlinecite{Yamashita03}] in an attempt to explain the specific heat
enhancement. The numerical calculations have been performed for the
supersymmetric case where all spin/orbital fluctuations are controlled by
the same nearly critical interaction parameter. In the numerical
calculations the instability is obtained only around the center of the
Brillouin zone and the large mass enhancement is due to contributions from
all possible spin/orbital fluctuations at ${\bf Q}$=0. In our approach we
take the complementary view point that (i) only the spin fluctuations are
close to be critical and (ii) the critical spin fluctuations at $T=0$ are
located in a wide region of ${\bf Q}$-space far away from the Brillouin
zone center in accord with the experimental evidence from inelastic neutron
scattering.  Thus in our model the extended region of nearly critical spin
fluctuations in momentum space is a promising candidate for the large mass
enhancement.


\section{Electronic band structure and  unenhanced dynamic spin
susceptibility of L\lowercase{i}V$_2$O$_4$}

\begin{figure}[tbhp!]
\includegraphics[width=0.45\textwidth]{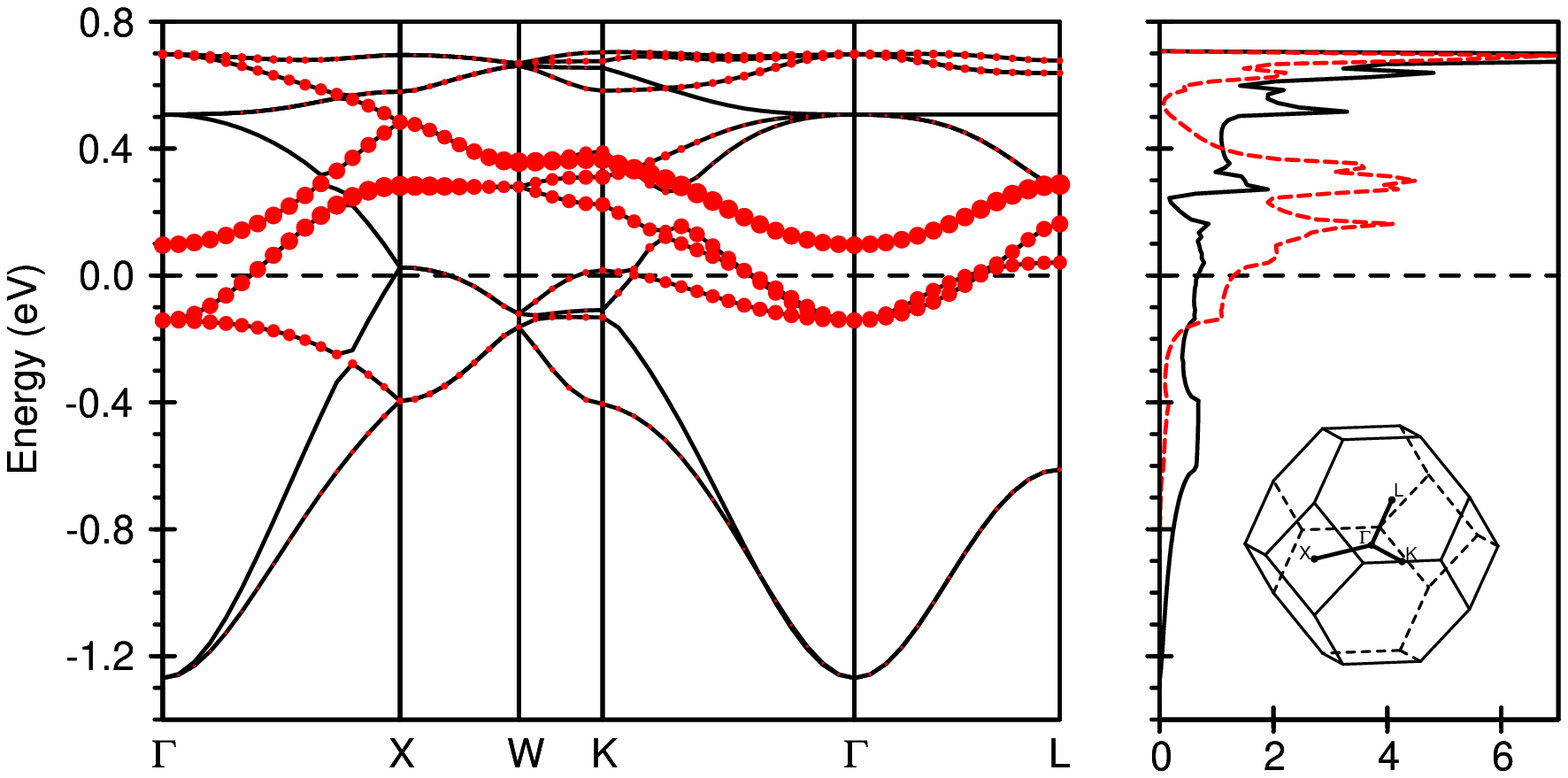}
\caption{\label{fig:ldabnd} (Color online) LDA bands originating from V
$d$ $a_{1g}$ and $e'_{g}$ states (left panel) and orbital resolved
densities of states states per eV, per V-atom (right panel). Densities of
$a_{1g}$ and $e'_{g}$ states are plotted by dashed and solid lines,
respectively. The size of circles in the left panel is proportional to the
weight of the $a_{1g}$ orbital in the state. The Fermi energy is at zero.}
\end{figure}

To define a reciprocal lattice for the cubic spinel structure of
LiV$_2$O$_4$, it is convenient first to introduce the orthogonal basis
vectors ${\bf G}^{(\alpha)}$, (${\alpha}=x,y,z$), of the length
$a^{\ast}={2\pi}/{a}\simeq 0.76 \AM$; here $a$=8.23\AA\ is the lattice
parameter.\cite{Kondo99} The irreducible Brillouin zone (BZ) of the
underlying fcc lattice is a polyhedron depicted in Fig.\ \ref{fig:ldabnd},
where X, Y, and Z points on their faces are given by the end points of the
vectors $\pm{\bf G^{(\alpha)}}$. A cube that encloses the polyhedron
provides us with a larger cubic BZ most appropriate to characterize
periodic properties of $\chi({\bf Q},\omega)$ in ${\bf Q}$-space. An
arbitrary wavevector ${\bf Q}$ will be denoted by ${\bf Q}={\bf q}+{\bf
G}$, where ${\bf q}$ belongs to the first cubic BZ and ${\bf G}$ is a
reciprocal lattice vector; ${\bf G}=2\sum_{\alpha}n_{\alpha}{\bf
G}^{(\alpha)}$, where $n_{\alpha}$ are integer.
  
In the present work, the band structure for LiV$_2$O$_4$ is calculated
within the LDA in the framework of LMTO (linear muffin tin orbitals) and
using the atomic sphere approximation. The main features of the calculated
band structure agree well in many details with the LDA results obtained by
other authors.\cite{Matsuno99,Singh99, Eyert99}

In LiV$_2$O$_4$ the vanadium 3$d$-bands and the oxygen 2$p$-bands are well
separated by an energy $\sim$2 eV. The octahedral component of the crystal
field is strong enough to split the vanadium 3$d$ bands into two separate
and non-overlapping bands originating from $t_{2g}$ and $e_g$ orbitals.  A
weak trigonal component of the crystal field produces a further splitting
of the low-lying set of $t_{2g}$ orbitals into ($a_{1g}+e^{\prime}_g$)
orbitals.  The latter dominate the electronic states near the vicinity of
the Fermi energy. The complexity of the band structure results in many
sheets of the Fermi surface.

Recalling that LiV$_2$O$_4$ is found to be a paramagnet down to very low
temperatures, we assume that the system remains spin disordered at all
temperatures, and carry out a calculation of the dynamic spin
susceptibility for $T=0$. Without external magnetic field the longitudinal
and the transverse susceptibility are the same.

In the multi-band model under consideration, the time- and space-Fourier
transform of the dynamic spin susceptibility for the LDA band electrons takes
the familiar form 
\begin{eqnarray}
\chi^{(0)}_{\bf G,G'}({\bf q},\omega)&=&\frac{1}{V}\sum_{\eta,\eta'}\sum_{\bf
k}\gamma_{\eta^{\prime} \eta}({\bf {k,q}; \omega})  \label{a1} \\
&&\times \left\langle{\bf k-q}/2,\eta |e^{-i\left({\bf q+G}\right){\bf r}}
|{\bf k+q}/2, \eta'\right\rangle \nonumber \\
&&\times \left\langle{\bf k+q}/2, \eta' | e^{i\left({\bf
q+G'}\right){\bf r}}|{\bf k-q}/2,\eta\right\rangle\,, \nonumber
\end{eqnarray} 
\begin{equation}
\gamma_{\eta^{\prime} \eta}({\bf k},{\bf q};\omega) = 
-\frac{f_{\eta'}({\bf k+q}/2)-f_{\eta}({\bf k-q}/2)}
{E_{\eta'}({\bf {k+q}/2})-E_{\eta}({\bf k-q}/2)+\hbar\omega +i\delta}\,. 
\label{a2}
\end{equation}
In (\ref{a1}) and (\ref{a2}), the plane-wave matrix elements are calculated
on the basis of Bloch functions $\psi_{{\bf k}\eta}({\bf r})$
with the band index $\eta$ and energy $E_{\eta}({\bf k})$. Here
$f_{\eta}({\bf k})$ is the Fermi distribution function.

In the next section, the exchange-correlation enhanced spin susceptibility
$\chi_{\bf G,G^{\prime}}({\bf q},\omega)$ is derived within the RPA. In
this derivation, ${\bf G }\not={\bf G^{\prime}}$ matrix elements of
$\chi^{(0)}_{\bf G,G^{\prime}}({\bf q},\omega)$ are involved, which lead to
the well known problem of large matrix inversion.  Since the magnetically
active 3$d$-orbitals of vanadium ions are rather compact, a large set of
the reciprocal lattice vectors ${\bf G}$ has to be taken into account and
thus a matrix $\chi^{(0)}_{\bf G,G^{\prime}}({\bf q},\omega)$
of high dimension has to be inverted. One encounters a similar problem when
calculating the inverse dielectric matrix for electrons in transition
metals.\cite{Hanke73} To overcome this difficulty, an elegant procedure
based on a simple and reliable approximation has been proposed and
developed by several authors.\cite{Hanke73,Callaway83} We are now in a
position to discuss some preliminaries of this procedure, the central point
of which is the search for a separable form for the plane-wave matrix
elements entering Eq.~(\ref{a1}).

Taking into account a four-site basis within the primitive lattice cell,
there are altogether twelve bands of dominant $d$-character which originate
from partially occupied $a_{1g}+e^{\prime}_g$ orbitals of the vanadium
ions. Therefore, one expects that the Bloch functions $\psi_{{\bf
k}\eta}({\bf r})$ for the actual $d$-bands near the Fermi energy can be
well described by the orbital basis set $\phi_{{\bf k },\bm{\tau}m}$ of
Bloch functions defined in terms of atomic-like
$a_{1g}+e^{\prime}_g$-orbitals localized on the pyrochlore lattice sites
${\bf j} + \bm{\tau}$:
\begin{equation}
\phi_{{\bf k },{\bm{\tau}}m}=\frac{1}{\sqrt{N}}
\sum_{{\bf j}}e^{i\bf kj}w_m(\bf r-j-\bm{\tau}). 
\label{a3}
\end{equation}
Here the sum is over $N$ ${\bf j}$-sites of the underlying fcc lattice. The
$\bm{\tau}$-vectors form a four-point basis. Furthermore, $w_m({\bf
r-j}-\bm{\tau})$ is the $m$-th orbital from the $a_{1g}+e^{\prime}_g$
orbital set on the pyrochlore lattice site ${\bf j}+\bm{\tau}$. For
shortening the notation, we use below a composite index $(\bm{\tau} m)$ to
refer to $w_m({\bf r-j}-\bm{\tau} )$ as the $(\bm{\tau} m)$-th orbital
belonging to the $\bf j$-th primitive lattice cell. We recall the following
unitary transformation between the band and the orbital representations for
Bloch functions
\begin{equation}
\psi_{{\bf k}\eta}({\bf r})= \sum_{\bm{\tau} m}
a_{(\bm{\tau}m)\eta}({\bf k}) \phi_{{\bf k},\bm{\tau} m}\,. 
\label{a4}
\end{equation}
Here $a_{(\bm{\tau}m)\eta}({\bf k})$ are elements of the
(12$\times$12)-matrix satisfying the following orthogonality and
completeness relations
\begin{eqnarray}
\sum_{\bm{\tau}m} 
a^{\ast}_{(\bm{\tau}m)\eta}({\bf k})a_{(\bm{\tau}m)\eta'}({\bf k}) &=&
\delta_{\eta  \eta'}\,,\nonumber \\  
\sum_{\eta} a_{(\bm{\tau}m)\eta} ({\bf k}) 
a^{\ast}_{(\bm{\tau}'m')\eta}({\bf k}) &=&
\delta_{\bm{\tau}m,\bm{\tau}'m'}\,,  
\label{a5}
\end{eqnarray}
where $\delta$ is the Kronecker symbol. The plane-wave matrix
elements in (\ref{a1}) can now be written as 
\begin{widetext}
\begin{equation}
\left\langle{\bf k+q}/2, \eta' | e^{i\left({\bf q+G'}\right){\bf r}}|{\bf
k-q}/2,\eta\right\rangle =
\sum_{\bm{\tau}'m'}\sum_{\bm{\tau}m}
A^{\eta' \eta}_{(\bm{\tau}^{\prime}m')(\bm{\tau}m)}({\bf k},{\bf q})
{\mathcal{F}}_{(\bm{\tau}'m')(\bm{\tau}m)}({\bf k},{\bf q+G'}), 
\label{a6}
\end{equation}
\end{widetext}
where 
\begin{equation}
A^{\eta' \eta}_{(\bm{\tau}'m')(\bm{\tau}m )}({\bf k},{\bf q}) =
a^{\ast}_{(\bm{\tau}'m')\eta'}({\bf k+q}/2)a_{(\bm{\tau}m)\eta}({\bf k-q}/2),
\label{a7}
\end{equation}
while the form-factor ${\mathcal{F}}$ involves integrals over pairs of
$w_m$-orbitals localized either at the same site or at two different
lattice sites. The compactness of these orbitals suggests to neglect the
two-center integrals, which leads to the following approximation for the
form factor
\begin{widetext}
\begin{eqnarray}
{\mathcal{F}}_{(\bm{\tau}'m' )(\bm{\tau}m )}({\bf k},{\bf q+G}) &\simeq&
\delta_{\bm{\tau}'\bm{\tau}} e^{i({\bf q+G})\bm{\tau}}
\int d{\bf r} w_{m'}({\bf r}) e^{i({\bf q+G}){\bf r}} w_{m}({\bf r}) 
\nonumber\\
&=& \delta_{\bm{\tau}'\bm{\tau}}F_{(\bm{\tau}m')(\bm{\tau}m)}({\bf q+G}). 
\label{a8}
\end{eqnarray}
\end{widetext}

We have checked that the neglect of the overlap integrals between the
calculated $d$ orbitals on different sites in LiV$_2$O$_4$ is a rather
accurate approximation. The approximation results, in particular, in a
${\bf k}$-independence of the form-factor $F$ in (\ref{a8}), the advantage
of which will be exploited in the next section. It is worth noting that in
(\ref{a8}) the factors $e^{i({\bf q+G})\bm{\tau}}$ are invariant under
translations by 2${\bf G}$ in reciprocal space.

For shortening the notation, it is helpful to replace in the upper
equations the double ``orbital'' index ${(\bm{\tau}'m')(\bm{\tau}m)}$ by a
symbol $L$ and to introduce the following orbitally projected expression
\begin{eqnarray}
\gamma_{LL^{\prime}}({\bf q},\omega) &=&
\frac{1}{N}\sum_{\eta,\eta'}\sum_{\bf k} 
\left[A^{\eta' \eta}_{L}({\bf k},{\bf q})\right]^{\ast} \nonumber \\
&&\times \gamma_{\eta^{\prime}\eta}({\bf {k,q};\omega})\ 
A^{\eta^{\prime} \eta}_{{L}^{\prime}}({\bf k},{\bf q})\,. 
\label{a9} 
\end{eqnarray}
In the following the left-hand side of Eq.~(\ref{a9}) is referred to as the
matrix $\gamma$ in the orbital $L$-representation.  The elements of the
orbital $\gamma$-matrix are periodic in reciprocal space:
$\gamma_{LL^{\prime}}({\bf q}+ {\bf G},\omega)=\gamma_{LL^{\prime}}({\bf
q},\omega)$. With these notations, the LDA spin susceptibility (\ref{a1})
takes the following form
\begin{equation}
\chi^{(0)}_{\bf G,G'}({\bf q},\omega) =
\sum_{LL'}F^{\ast}_L({\bf q+G}) \gamma_{LL'}({\bf q},\omega) 
F_{L'}({\bf q+G'}).      
\label{a10}
\end{equation}
It is apparent from Eqs.\ (\ref{a2}), (\ref{a9}), and (\ref{a10}) that the
computation of three types of quantities is required in a wide range of
${\bf q}$, ${\bf k}$, and $\omega$: these are the multi-band matrix
$\gamma_{\eta^{\prime} \eta}$, the matrix elements $A^{\eta' \eta}_{L}$ and
the form factor $F_L$.

For an analysis of inelastic neutron scattering as a function of momentum
$\hbar ({\bf q}+{\bf G})=\hbar{\bf Q}$ and energy $\hbar \omega$ transfer,
the diagonal (${\bf G}^{\prime}={\bf G}$) term $\chi_{\bf G,G}({\bf
q},\omega)=\chi({\bf Q},\omega)$ should be calculated including the
relevant electron interactions. This will be discussed in the next
section. At this stage, we calculate first the unenhanced $\chi^{(0)}({\bf
Q},\omega)$. This is done along three high-symmetry directions $\Gamma$X,
$\Gamma$K and $\Gamma$L in reciprocal space which are further abbreviated
by X, K and L, respectively. For each of these directions D, wavevectors
${\bf q}$ and ${\bf G}$ are parallel and the use of a modulus $Q^D$ of the
wavevector ${\bf Q}={\bf q}+{\bf G}$ is sufficient, provided the system has
space inversion symmetry.
 
\begin{figure}[tbhp!]
\includegraphics[width=0.35\textwidth]{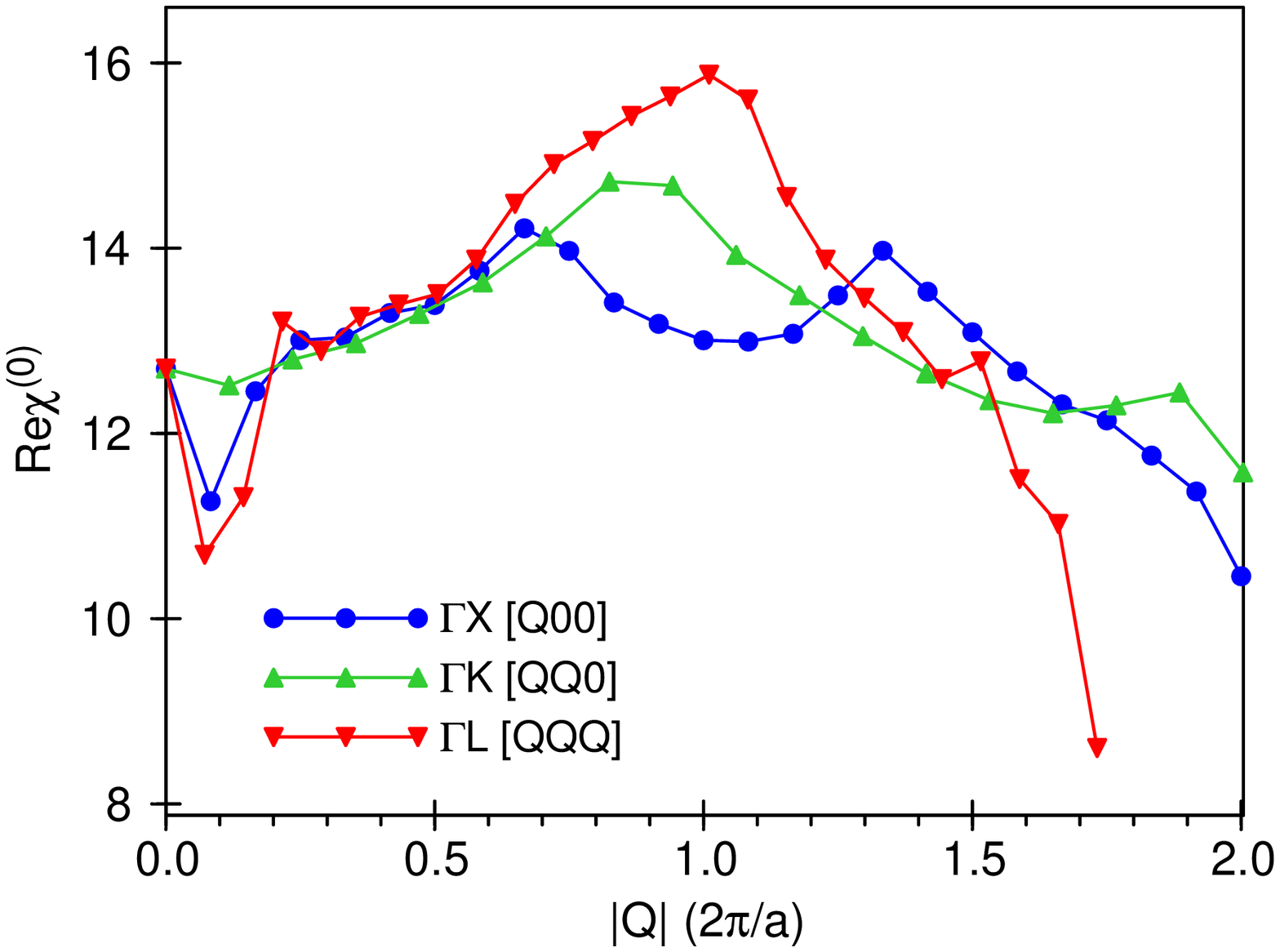}
\caption{\label{fig:xi0q} (Color online) Unenhanced static spin
susceptibility $\chi^{(0)}({\bf Q},0)$, in states per eV, per primitive
cell (4 V-atoms), calculated along high symmetry directions as a function
of $|{\bf Q}|$.}
\end{figure}

In the static ($\omega =0$) limit, the spin susceptibility $\chi^{(0)}({\bf
Q},0)$ is calculated and the results are displayed in Fig.\ \ref{fig:xi0q}
as a set of functions $\chi^{(0)}_D(Q)$ with wavevectors of length $Q$
along the directions D = X, K, L. For $Q>$2$a^{\star}$ the susceptibility
is considerably suppressed due to the form factor.

One notices a significant variation of $\chi^{(0)}_D(Q)$ over the whole
range of the wavevectors chosen. Since we intent to apply the RPA theory,
the positions of the maxima in ${\bf Q}$-space of the unenhanced
susceptibility $\chi^{(0)}({\bf Q},0)$ provide a valuable information: at
these wavevectors and in their vicinity one expects the strongest spin
correlations when the enhanced $\chi({\bf Q},0)$ is calculated.
 
For each direction D, two maxima of $\chi^{(0)}_D(Q)$ at
wavevectors $Q_{c1}^D$ and $Q_{c2}^D$ are found; the second strongly
suppressed maximum at $Q_{c2}^{L}>$2$a^{\star}$ along the L-direction is
not depicted here. The maxima at smaller $Q_{c1}^D$ ($<Q_{c2}^D$) are
located in the first cubic BZ while those at $Q_{c2}^D$ are in the next BZ.

From now on let us refer to the three wavevectors $Q_c=Q_{c1}^{X,K,L}$ as
the ``critical'' ones. From Fig.\ \ref{fig:xi0q} we obtain the following
estimates: $Q_{c1}^X\approx 0.50\AM$, $Q_{c1}^K\approx 0.65\AM$, and
$Q_{c1}^L\approx 0.75\AM$. Remarkably, these three values occur within the
range 0.4$\AM \lesssim Q\lesssim 0.8\AM$ where the main quasielastic
neutron scattering is observed.\cite{Krimmel99,Lee01,Murani04}

\begin{figure}[htbp!]
\includegraphics[width=0.35\textwidth]{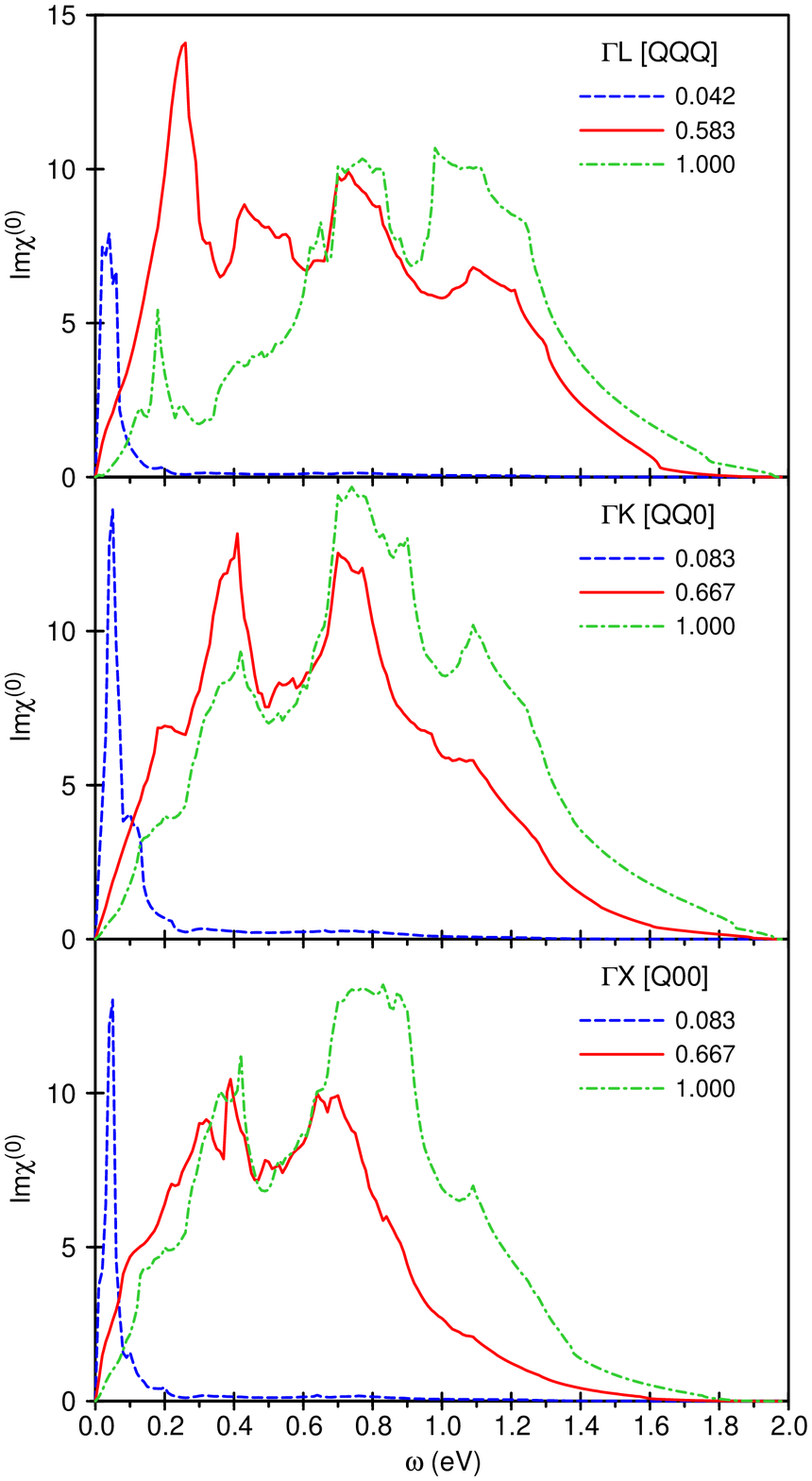}
\caption{\label{fig:xi0w} (Color online) Unenhanced Im$\chi^{(0)}({\bf
Q},\omega )$ calculated at three different wave-vectors chosen along the
high symmetry directions D=L, K, X.  The ``critical'' wave-vectors
$Q^{L,K,X}_{c1}$ are denoted by the Cartesian component lenghts: 0.583;
0.667; 0.667 (in units $a^{\star}$), respectively.}
\end{figure}

The frequency distribution of spin fluctuations of the system of
independent electrons may be seen from Fig.\ \ref{fig:xi0w} where
representative results of the calculated Im$\chi^{(0)}_D({\bf Q},\omega)$
are shown for three wavevectors of different length chosen along each of
the high-symmetry directions D. For instance, at small wavevectors $Q$
along the L-direction, Im$\chi^{(0)}_L(Q,\omega)$ shows a low-$\omega$
single-peaked structure. With increasing values of $Q$, the low-$\omega$
peak moves upward and an additional broad distribution arises and grows
gradually at higher frequencies. At the ``critical'' wavevector $Q_{c1}^L$,
the pronounced low-$\omega$ peak at $\hbar \omega \sim$0.2 eV still
survives. However, a large portion of the spectral weight is found now
within a broad distribution over much higher frequencies.  As $Q$ grows
further, the low-$\omega$ peaked feature gets suppressed and most of the
spectral weight is shifted to the high-frequency region. As long as
considerable spectral weight of the low-$\omega$ peaked feature of
Im$\chi^{(0)}_L(Q,\omega)$ is present, like at $Q\approx Q_{c1}^L$, its
position on the $\omega$-axis is regarded as a bare (unrenormalized) spin
relaxation rate $\Gamma^{(0)}(Q)$ characteristic of low-$\omega$ spin
fluctuations. As expected, bare values of $\Gamma^{(0)}(Q)$, for instance,
$\Gamma^{(0)}(Q_{c1}^L)\approx$0.2 eV, are much too high to explain the
slow spin dynamics found in the experiment.\cite{Krimmel99,Lee01,Murani04}
Similar arguments are also applicable to the behavior of
Im$\chi^{(0)}_D(Q,\omega )$ along K- and X-directions displayed in Fig.\
\ref{fig:xi0w}.  Therefore interactions between quasiparticles must play an
essential role to reach the observed energy scale ${\Gamma}$ of spin
fluctuations.  In the next section, interaction effects are taken into
account in the form of RPA theory.


\section{RPA approach to the dynamic spin susceptibility of 
L\lowercase{i}V$_2$O$_4$}

Within linear response theory, the time-Fourier transform of the dynamic
spin susceptibility obeys the following integral equation
\begin{eqnarray}
\chi({\bf r},{\bf r'};\omega)&=&\chi^{(0)}({\bf r},{\bf r'};\omega) 
\label{a11} \\ 
&&+ \int d{\bf r''} \chi^{(0)} ({\bf r},{\bf r''};\omega) 
\mathcal{K}({\bf r''})\chi({\bf r''},{\bf r'};\omega). \nonumber
\end{eqnarray}
Here $\mathcal{K}({\bf r})$ is a first derivative of the spin-dependent
part of the exchange-correlation potential $V^{s}_{xc}$ taken in the local
approximation.  Since $\mathcal{K}({\bf r})$ is a periodic function,
$\mathcal{K}({\bf r +j})=\mathcal{K}({\bf r})$, its reciprocal-space
counterpart $\mathcal{K}({\bf G})$ is a function of the reciprocal-lattice
vectors ${\bf G}$. After performing the double space-Fourier
transformation, Eq.~(\ref{a11}) is expressed as
\begin{eqnarray}
\chi_{{\bf G}{\bf G}^{\prime}}({\bf q},\omega) &=& 
\chi^{(0)}_{{\bf G}{\bf G}^{\prime}}({\bf q},\omega) \label{a12} \\
&&+ \sum_{{\bf G}_1{\bf	G}^{\prime}_1} 
\chi^{(0)}_{{\bf G}{\bf G}_1} ({\bf q},\omega)
\mathcal{K}_{{\bf G}_1{\bf G}^{\prime}_1} 
\chi_{{\bf G}^{\prime}_1{\bf G}^{\prime}}({\bf q},\omega), \nonumber
\end{eqnarray}
with $\mathcal{K}_{{\bf G}_1{\bf G}^{\prime}_1}=\mathcal{K}({\bf G}_1-{\bf
G}^{\prime}_1)$. Solving of Eq.~(\ref{a12}) requires an inversion of an
infinite matrix $(1-\chi^{(0)}\mathcal{K})_{{\bf G}{\bf G}^{\prime}}$ in
reciprocal-lattice $({\bf G})$ representation. To tackle the problem, first
one has to examine the convergence of the matrix elements of the
susceptibility as the dimension of the $({\bf G})$-basis increases. To
avoid this we use below a procedure suggested and developed in Refs.\
[\onlinecite{Hanke73,Callaway83}].

The separable form (\ref{a10}) of $\chi^{(0)}=F^{\ast}\gamma F$ allows us
to solve the matrix inversion problem by transforming it from $\bf
G$-representation to the orbital $L$-representation. Actually, the
iteration procedure applied to Eq.~(\ref{a12}) yields the matrix expansion
$\chi=F^{\ast}\gamma (1+M+M^2+\dots)F$, where the matrix elements of $M$
are
\begin{eqnarray}
M_{L_1L_3}({\bf q},\omega)&=&\sum_{L_2}\left[\sum_{{\bf G}{\bf G}^{\prime}} 
F_{L_1}({\bf q+G})\mathcal{K}_{{\bf G}{\bf G}^{\prime}} 
F_{L_2}^{\ast}({\bf q+G'})\right] \nonumber \\
&&\times \gamma_{L_2L_3}({\bf q},\omega)\,.
\label{a13}
\end{eqnarray}
Assuming a convergence of the above power series of $M$ at any ${\bf q}$
and $\omega$, it can be converted into a familiar form
$\left[1-M\right]^{-1}$. Before doing this, let us consider the matrix
elements $\left[ \sum F{\mathcal{K}}F^{\ast}\right]_{L_1L_2} $ entering the
definition (\ref{a13}) of the matrix $M$ and specify its indices as
$L_1={(\bm{\tau}_1m^{\prime}_1)(\bm{\tau}_1m_1)}$ and
$L_2={(\bm{\tau}_2m^{\prime}_2)(\bm{\tau}_2m_2)}$. Then by using the
definition (\ref{a8}) for the approximate form factor $F$ and the
inverse Fourier transformation for $\mathcal{K}$, a matrix element
$\left[\sum F{\mathcal{K}}F^{\ast}\right]_{L_1L_2} $ can be presented as a
space integral of a product function of $\mathcal{K}({\bf r})$ and two
pairs of localized orbitals $w_m({\bf r-j}-\bm{\tau})$. In general, there
are one-center and two-center integrals. We neglect interactions between
different sites and retain only the one-center integrals. Thus we use the
approximation
\begin{widetext}
\begin{eqnarray}
\left[ \sum_{{\bf G}{\bf G}^{\prime}} F({\bf q+G})
{\mathcal{K}}_{{\bf G}{\bf G}^{\prime}} F^{\ast}({\bf
q+G'})\right]_{L_1L_2} \simeq 
\delta_{\bm{\tau}_1\bm{\tau}_2}
{\mathcal{K}}(m_1^{\prime} m_1;m_2^{\prime} m_2)\,, \nonumber \\
{\mathcal{K}}(m^{\prime}_1 m_1; m^{\prime}_2 m_2)= 
 v_0  \int\limits_{v_0} d{\bf r} 
w_{m_{1}^{\prime}}({\bf r}) w_{m_{1}}({\bf r}) 
{\mathcal{K}}({\bf r})
w_{m_{2}^{\prime}}({\bf r}) w_{m_2}({\bf r})\,.
\label{a14}
\end{eqnarray}
\end{widetext}
Furthermore, we neglect Hund's rule correlations and keep only the diagonal
matrix elements $\mathcal{K}(m,m;m,m)$. Thus, the $m$-independent unique
coupling constant ${\mathcal{K}}$ is the only adjustable parameter in our
approximate approach. Finally, we arrive at the following expression for
the enhanced spin susceptibility
\begin{widetext}
\begin{eqnarray}
\chi_{\bf G,G^{\prime}}({\bf q},\omega) =
\sum_{LL'L''}F^{\ast}_L({\bf q+G})\gamma_{LL''}({\bf q},\omega) 
\left(\left[1-{\mathcal{K}}\gamma({\bf q},\omega)\right]^{-1}\right)_{L''L'} 
F_{L'}({\bf q+G'}),     
\label{a15}
\end{eqnarray}
\end{widetext}
where the elements of the matrix $\gamma({\bf q}, \omega)$ are given by
(\ref{a9}). The essence of the procedure is instead of inverting the large
matrix needed to solve the Eq.~(\ref{a12}) to invert the smaller matrix
$\left(1-\chi^{(0)}{\mathcal{K}}\right)_{LL^{\prime}}$ in the orbital
basis.

Because of the composite character of orbital indices $L$, matrix
operations in $L$-representation require additional comments. In
particular, matrix elements of a transposed matrix $\tilde{\gamma}$ are
related to those of the direct matrix as
$\tilde{\gamma}_{L_1L_2}=\gamma_{\bar{L}_2\bar{L}_1}$, where each composite
index $L_n={(\bm{\tau}^{\prime}_nm^{\prime}_n)(\bm{\tau}_nm_n)}$ is also
transposed to give
$\bar{L}_n={(\bm{\tau}_nm_n)(\bm{\tau}^{\prime}_nm^{\prime}_n)}$. Therefore,
for a Hermitian conjugate matrix $\gamma^{\dagger}$ one has
$\gamma^{\dagger}_{L_1L_2}=\gamma^{\ast}_{\bar{L}_2\bar{L}_1}$. There are
144 distinct indices $L$, therfore $\gamma$ is a
(144$\times$144)-matrix. The large number of elements
$\gamma_{LL^{\prime}}({\bf q},\omega)$ that have to be determined for each
value of ${\bf q}$ and $\omega$ makes the numerical calculations of $\chi$
a rather complicated problem.  At some stage pointed out below, we will
introduce a simplification (``diagonal'' approximation for the matrix
$\gamma$) allowing us to reduce the matrix dimensionality and, thus, make a
numerical analysis feasible.

According to the definition (\ref{a9}), the matrix $\gamma\left({\bf q},
\omega\right)$ can be generally decomposed as follows 
\begin{equation}
\gamma({\bf q},\omega) = \gamma^{(1)}({\bf q},\omega)
+i\gamma^{(2)}({\bf q},\omega)\,.
\label{a16}
\end{equation}
Here, the elements of the Hermitian matrix
$\gamma^{(1)}=(\gamma+\gamma{\dagger})$/2 are even functions of $\omega$
while those of the anti-Hermitian matrix
$\gamma^{(2)}=(\gamma-\gamma{\dagger})$/2$i$ are odd functions of
$\omega$. Note that $\gamma^{(2)}({\bf q},0)$=0 for all ${\bf q}$.

Let us consider the static, $\omega=0$, limit and solve the eigenvalue
problem for the Hermitian matrix $\gamma({\bf
q},0)=\gamma^{(1)}({\bf q},0)$:
\begin{equation}
\sum_{L^{\prime}} \gamma_{LL^{\prime}}({\bf q},0)
V^{(k)}_{L^{\prime}}({\bf q})=\lambda^{(k)}({\bf q})V^{(k)}_L({\bf q})\,,  
\label{a17}
\end{equation}
where $V^{(k)}_L $ are components of the $k$-th eigenvector with the
eigenvalue $\lambda^{(k)}$. The static spin susceptibility takes the
following form
\begin{equation}
\chi_{\bf G,G}({\bf q}, 0) =
\sum_k\left|{\mathcal{F}}^{(k)}({\bf q},{\bf G})\right|^2
\frac{\lambda^{(k)}({\bf q})} {1-{\mathcal{K}}\lambda^{(k)}({\bf q})}\,,
\label{a18}
\end{equation}
with ${\mathcal{F}}^{(k)}({\bf q},{\bf G})=\sum_{L}F^{\ast}_L({\bf
q+G})V^{(k)}_L({\bf q})$. In solving Eq.~(\ref{a17}), we adopt a
``diagonal'' approximation \cite{Callaway83} which consists in retaining in
$\gamma_{LL^{\prime}}$ only the diagonal composite indices, $L={(\bm{\tau}m
)(\bm{\tau}m)}$ and $L^{\prime}={(\bm{\tau}^{\prime}m^{\prime}
)(\bm{\tau}^{\prime}m^{\prime})}$. Then $\gamma_{LL^{\prime}}$ becomes a
(12$\times$12)-matrix. A partial justification is that the moduli of the
off-diagonal ${(\bm{\tau}^{\prime}m^{\prime} )\ne(\bm{\tau}m)}$, matrix
elements (\ref{a7}) are small in our calculations. To assess the accuracy
of the "diagonal" approximation, a small set of arbitrary wavevectors ${\bf
q}$ was chosen and Im$\chi^{(0)}({\bf q},\omega)$ was calculated (at ${\bf
G}={\bf G}^{\prime}=0$) as a function of $\omega$ including either all or
only "diagonal" elements of the matrix $\gamma$.  At each ${\bf q}$ probed,
we found rather tiny differences between the two curves.

Like in section II, we consider below only the high-symmetry X-, K- and
L-directions in reciprocal space. For each direction D, let us pick out the
``critical'' wavevector $Q_{c}^D$ and rearrange the set of
$\lambda^{(k)}(Q_{c}^D)=\lambda^{(k)}_{D}$ in the following order:
$\lambda_D^{(1)} >\lambda^{(2)}_D >\lambda^{(3)}_D >\dots$.  For brevity, a
set of nondegenerate $\lambda_D^{(k)}$ is discussed here. The first members
of the set $\lambda_D^{(k)}$ at $k=1,2,\dots$ are positive in order to
ensure $\chi_{\bf G,G}({\bf q}, 0)>0$, provided ${\mathcal{K}}$ is below
some critical value ${\mathcal{K}}_c$. It is apparent from (\ref{a18}) that
along $D$-direction the $\lambda_D^{(1)}$-mode is the most ``critical'' one
in the sense that $0\leq1-{\mathcal{K}}\lambda_D^{(1)}
<1-{\mathcal{K}}\lambda^{(2)}_D <1-{\mathcal{K}}\lambda^{(3)}_D <\dots$. A
magnetic instability occurs if the condition
$1-{\mathcal{K}}\lambda_D^{(1)}=0$ is fulfilled. This defines
${\mathcal{K}}_c$. Among the three directions under consideration, the
smallest ${\mathcal{K}}_c=$0.49 eV occurs along the L-direction at $
Q_{c1}^L\approx 0.75\AM$. Since for X- and K-directions the instability
conditions are fulfilled at values of ${\mathcal{K}}$ that are close, but
somewhat larger than ${\mathcal{K}}_c=$0.49 eV, the latter value is chosen
to be the best estimate of the critical value of the exchange-correlation
parameter ${\mathcal{K}}$. In LiV$_2$O$_4$, spin dynamics of a magnetically
disordered state with strongly enhanced spin fluctuations may be described
by Eq.~(\ref{a15}) with the parameter ${\mathcal{K}}$ approaching
${\mathcal{K}}_c$ from below.

\begin{figure}[htbp!]
\includegraphics[width=0.35\textwidth]{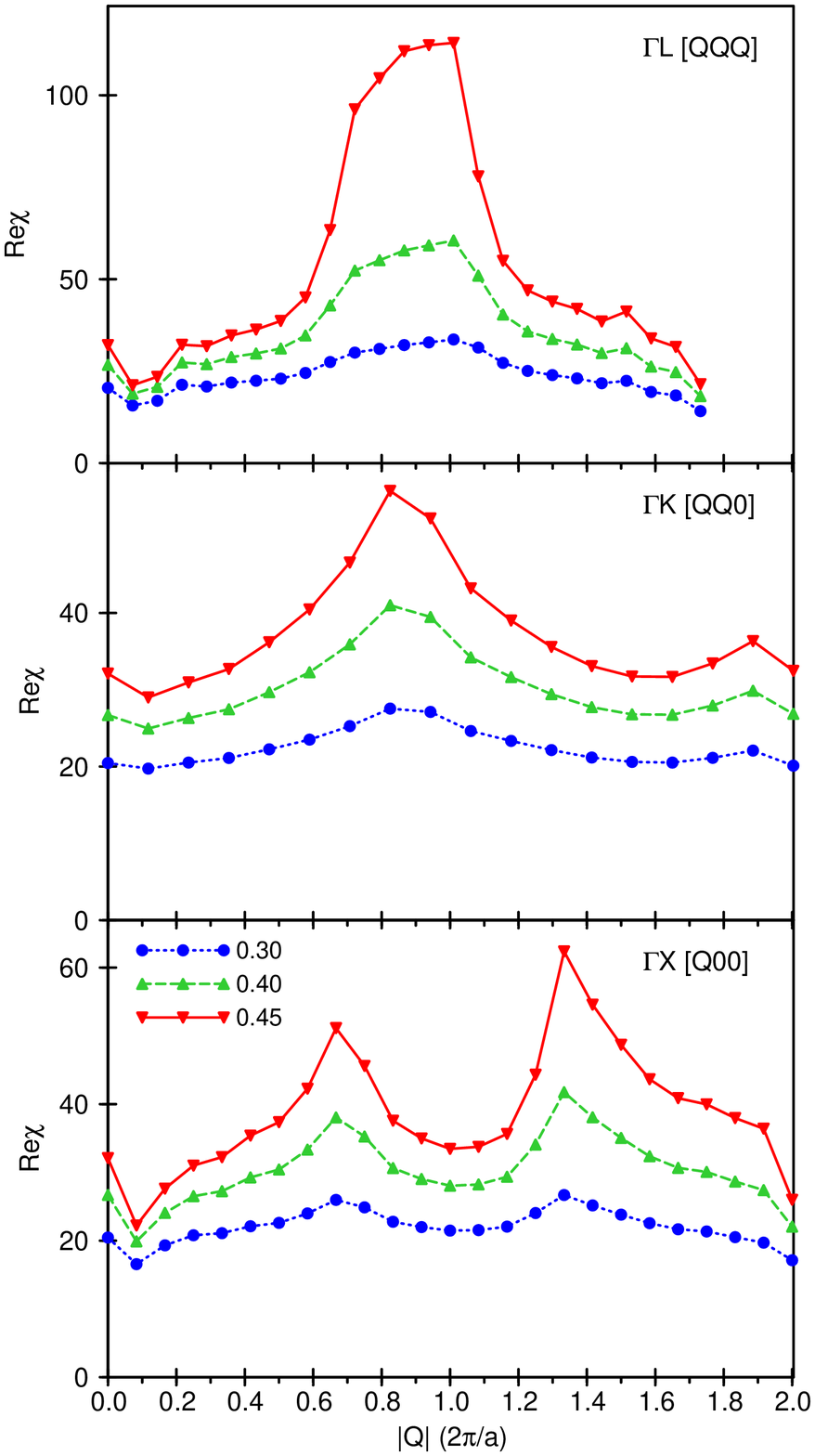}
\caption{\label{fig:xiq} (Color online) Evolution of the static spin
susceptibility $\chi ({\bf Q}, 0 )$ for different values of the
coupling constant ${\mathcal{K}}\to {\mathcal{K}}_c=$0.49eV. Representative
values of ${\mathcal{K}}$ (in eV) chosen are shown in the low panel.}
\end{figure}

The enhancement of the static susceptibility (\ref{a18}) in a spin
disordered state is controlled by a generalized Stoner factor
$S_{\mathcal{K}}({\bf Q})=\left[1-{\mathcal{K}} \lambda^{(1)}({\bf
Q})\right]^{-1}$, which is a function of both ${\mathcal{K}}$ and ${\bf
Q}$, with the obvious property $S_{{\mathcal{K}}}(Q_{c1}^L) \to \infty$ as
${\mathcal{K}}\to {\mathcal{K}}_c$. Thus the Stoner factor describes the
proximity of the system to the instability. As may be seen from Fig.\
\ref{fig:xiq}, the strongest enhancement of the static susceptibility
(\ref{a18}) is found for spin fluctuations at and nearby the ``critical''
wavevectors $Q_c$ and at $Q_{c2}^X$. At the same time, the Stoner factor is
of a moderate size for the remaining wavevectors.

Very similar  $Q$-dependence of the exchange-correlation effects is
seen from the calculated  imaginary part of the enhanced dynamic spin
susceptibility Im$\chi_D (Q,\omega)$.
 
\begin{figure}[htbp!]
\includegraphics[width=0.35\textwidth]{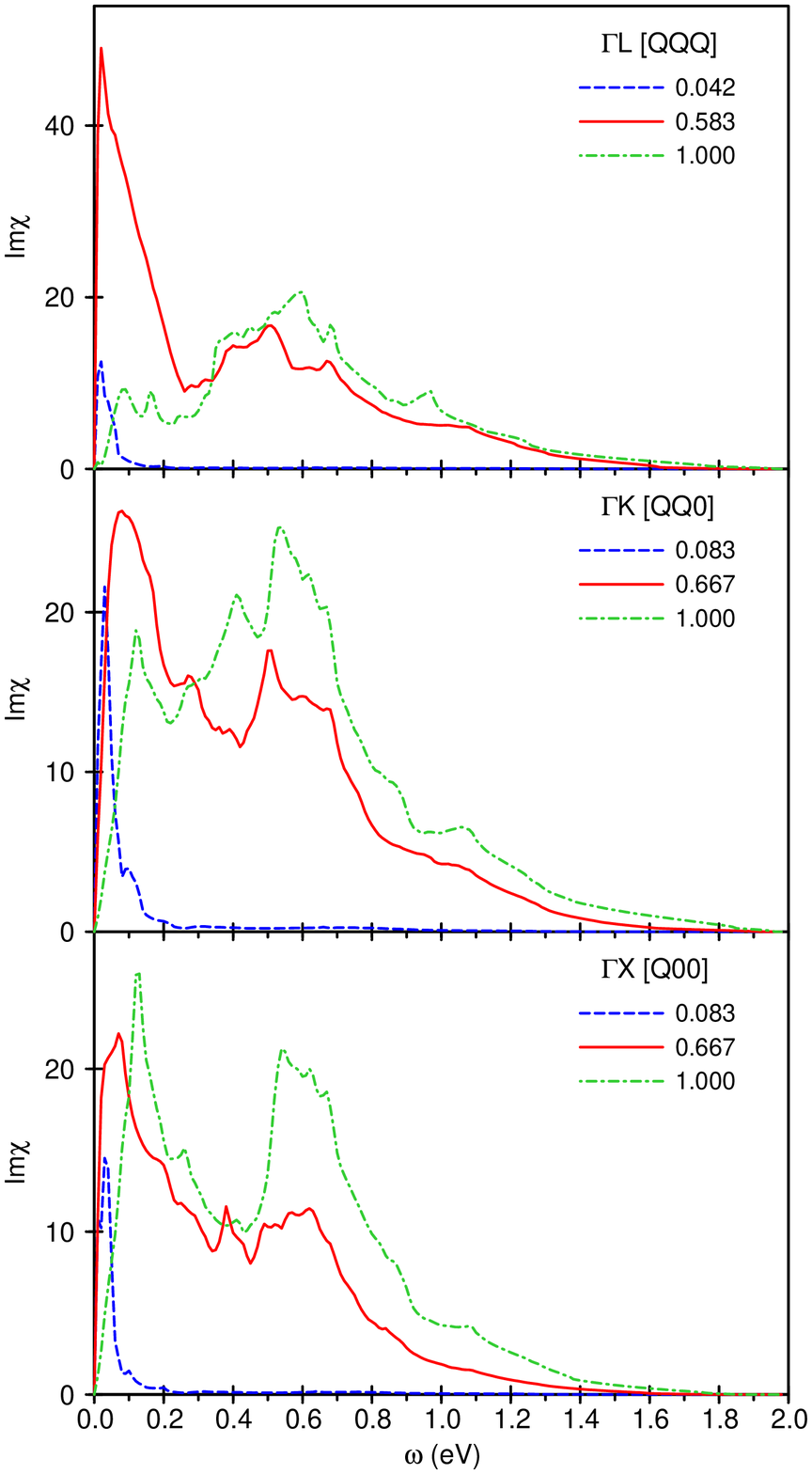}
\caption{\label{fig:xiw} (Color online) Enhanced Im$\chi({\bf Q},\omega)$
calculated at ${\mathcal{K}}= 0.95{\mathcal{K}}_c$ for the same wavevectors
as in Fig.\ \ref{fig:xi0w}. Note: the size along the vertical axis is
amplified to that in Fig.\ \ref{fig:xi0w}.}
\end{figure}

Results for the calculated Im$\chi_D(Q,\omega)$ are shown in Fig.\
\ref{fig:xiw}. They have to be compared with the unenhanced spin
susceptibility Im$\chi^{(0)}_D(Q,\omega)$ plotted in Fig.\
\ref{fig:xi0w}. Let us follow the evolution of the $\omega$-dependence of
Im$\chi_L(Q,\omega)$ along the L-direction with increasing
${\mathcal{K}}\to {\mathcal{K}}_c$. Like in Fig.\ \ref{fig:xi0w}, three
characteristic wavevectors are chosen to distinguish the behavior at
``critical'' $Q^L\simeq Q_{c1}^L $ from those at smaller and larger
wavevectors. As expected, the net frequency dependence of Im$\chi_L
(Q,\omega )$ is most strongly affected at $Q_{c1}^L$. In the upper panel of
Fig.\ \ref{fig:xiw} a large portion of the spectral weight is transfered to
the low energy region though at $Q_{c1}^L $ the overall spectral
distribution still extends up to $\sim$2 eV. The pronounced low-$\omega$
feature seen at $ Q_{c1}^L $ strongly dominates those at smaller and larger
wavevectors.

Generally, in the low-$\omega$ limit the imaginary part of $\chi(Q,\omega)$
can be accurately approximated by
\begin{equation}
\mathrm{Im}\chi (Q,\omega) \simeq z_Q\chi(Q)\omega/\Gamma({Q}),
\label{a18b}
\end{equation}
where the weight factor $z_Q <1$. Again, by choosing the L-direction as the
most representative one, we have compared the initial gradients of several
curves representing Im$\chi_L (Q_{c1},\omega)$ for different
${\mathcal{K}}\lesssim {\mathcal{K}}_c$, and found that the initial slope
is proportional to $S_{{\mathcal{K}}}^2(Q_{c1}^L)$ and thus grows very fast
as ${\mathcal{K}} \to {\mathcal{K}}_c$.  The factor
$S_{{\mathcal{K}}}^2(Q)$ stems from the Stoner enhanced static
susceptibility, $\chi(Q)\simeq S_{{\mathcal{K}}}(Q)\chi^{(0)}(Q)$, and a
renormalization of the spin relaxation rate, $\Gamma(Q) \simeq
\Gamma^{(0)}(Q)/S_{\mathcal{K}}(Q)$. Far away from $ Q_{c1}^L$ the same
relation holds, however, with much weaker enhancement factor,
$S_{{\mathcal{K}}}^2(Q^L)\ll S_{{\mathcal{K}}}^2(Q_{c1}^L)$. Concerning the
evolution of Im$\chi_D (Q,\omega )$ for D = X,K with increasing
${\mathcal{K}}$, our calculations show a similar behavior, i.e., strongly
enhanced spin fluctuations at $Q^D\sim Q_{c}^D$ in the low-$\omega$ region.
 
If the value of the adjustable coupling parameter ${\mathcal{K}}$ is tuned
close to ${\mathcal{K}}_c$ then $\Gamma(Q_c)$ may be reduced to values as
low as the experimentally observed ones.  However, this should not be taken
as a literal explanation of experimental data because close to the critical
condition RPA becomes unreliable. Furthermore, in our approach we ignored
the quantitative contribution of inter-site charge fluctuations to the
reduction of $\Gamma^{(0)}$.  Instead, our purpose is to show that like in
many other materials, in the case of LiV$_2$O$_4$ the RPA theory is a
useful tool to take into account the qualitative effects of strong electron
correlations which provide a basic mechanism for a strong reduction of spin
relaxation rate in a broad region of ${\bf Q}$-space.
 

\section{Discussion}

In this section we want to put our results  in a more  general context. 

Quasielastic neutron scattering studies of powder samples of LiV$_2$O$_4$
suggest that in the low-$T$ limit and in a range of wavevectors 0.4$\AM
\lesssim Q\lesssim 0.8\AM$ the vanadium spin system exhibits strongly
enhanced and slow spin fluctuations. A complete theoretical analysis of
these data would require calculations of the dynamic spin susceptibility in
a wide range of (${\bf Q}, \omega$)-space.  In principle, that is possible
within our approximation scheme except for the large computational time
required. The calculation of $\chi({\bf Q},\omega)$ can be performed in any
domain of (${\bf Q},\omega$)-space, not only along the high-symmetry ${\bf
Q}$-directions. However, even the limited results of sections II and III
provide a valuable and essential piece of information on the spin
fluctuation dynamics in LiV$_2$O$_4$.

Actually, a particular ${\bf Q}$-dependence of the calculated unenhanced
spin susceptibility $\chi^{(0)}({\bf Q},0)$ which involves the actual band
structure of LiV$_2$O$_4$, clearly displays in Fig.\ \ref{fig:xi0q} that
the particular, ``critical'', wavevectors $Q_{c1}^{X,K,L}$ are grouped in
the range 0.5$\AM \lesssim Q_c\lesssim 0.8\AM$ where the pronounced
quasielastic neutron scattering is observed. However, the bare spin
relaxation rate $\Gamma^{(0)}(Q)\sim$200 meV is much too large to be
compatible with experimental observations.  This disagreement has to be
attributed to strong electron correlations not included in LDA
calculations. Taking them into account by means of a RPA we have obtained
an enhanced susceptibility $\chi({\bf Q},\omega)$ as is seen by comparing
Figs.\ \ref{fig:xi0w} and \ref{fig:xiw}. As the coupling tends to the
critical value, ${\mathcal{K}}\rightarrow{\mathcal{K}}_c$, spin
fluctuations near $Q_c$ are strongly renormalized: a large fraction of the
spectral weight is shifted to low frequencies, thus resulting in an
enhanced low-energy response.

\begin{figure}[htbp!]
\includegraphics[width=0.4\textwidth]{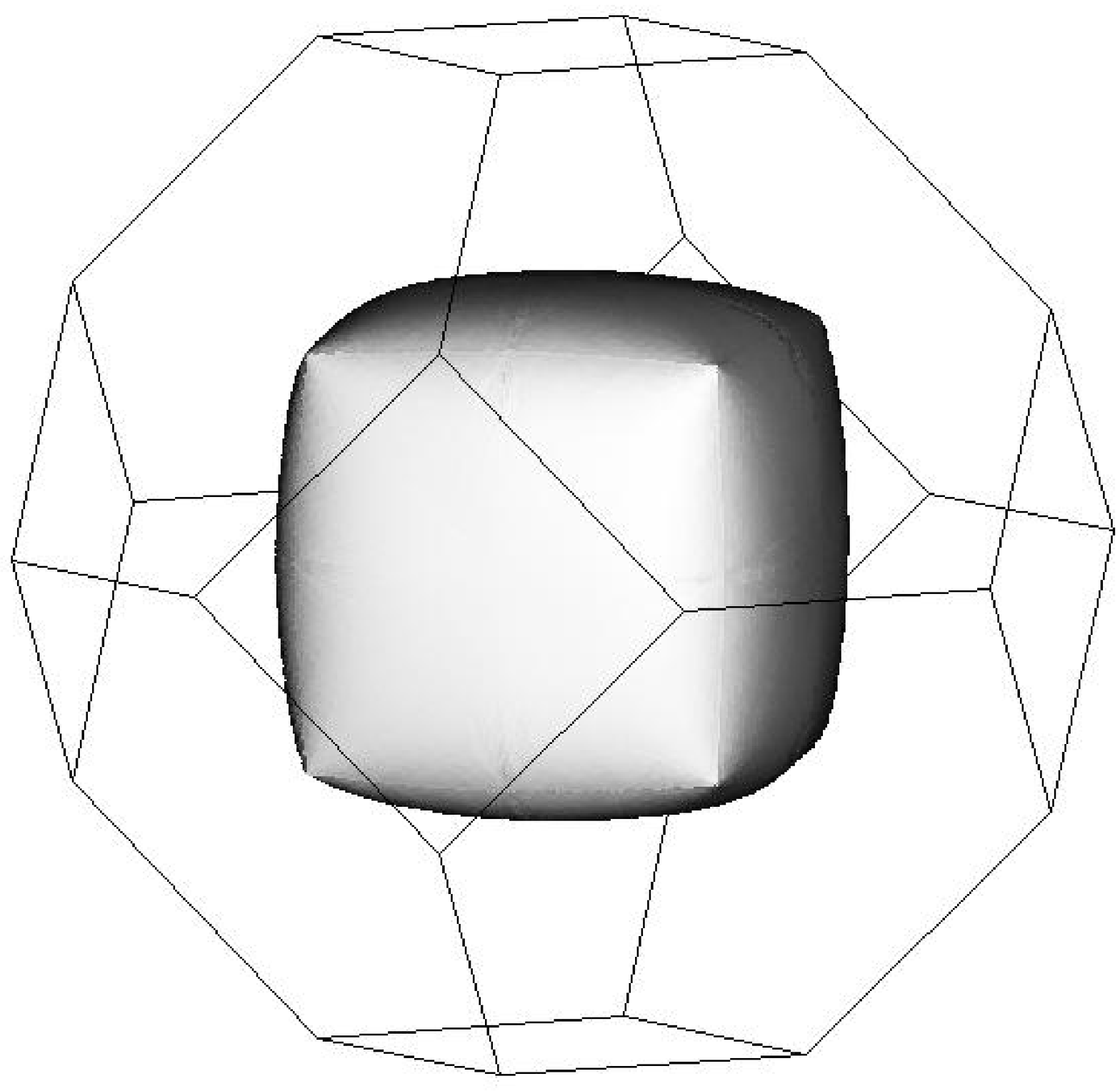}
\caption{\label{fig:qsurf} A surface in ${\bf Q}$-space representing the
lower bound for the ``critical'' region of strongly enhanced slow spin
fluctuations.  The upper bound is the larger congruent surface at a
distance $|\delta{\bf Q}|\simeq 0.3a^{\star}$ from the smaller one
shown.}
\end{figure}

Next we want to discuss in more detail the part of ${\bf Q}$-space with
slow spin fluctuations. The wavevectors ${\bf Q}_c = {\bf Q}_{c1}^{X,K,Z}$
are positioned on a smooth three dimensional surface to which we refer as
``critical'' ${\bf q}_c$-surface. This surface is in the first cubic
Brillouin zone and we shall replace ${\bf Q}$ by ${\bf q}$ in that case. We
have checked by additional calculations that spin fluctuations are slow
everywhere on the ${\bf q}_c$-surface. Moreover, the broad maxima of
$\chi({\bf q},0)$ in Fig.\ \ref{fig:xiq} suggest that the region of slow
fluctuations extends over a width of $(|{\bf q}_c|-\delta q/2)<|{\bf
q}|<(|{\bf q}_c|+\delta q/2)$, while the ${\bf q}_c$-surface lies midway
between two bounds. The lower bound of this region is shown by the surface
depicted in Fig.\ \ref{fig:qsurf}. It intersects the X-axis (and the
equivalent Y- and Z-axes) at $(q_{c1}^{X} -\delta q/2) \approx
0.5a^{\star}\approx0.4\AM\equiv q_1^{X}$. The upper bound is a congruent
surface crossing the X-axis at $(q_{c1}^{X} +\delta q/2) \approx
0.8a^{\star}\approx0.6\AM\equiv q_2^{X}$. Thus $q_1^{X}$ and $q_2^{X}$ are
minimal radii of two bounds. The width $\delta q$ is estimated to be
$\delta q\approx 0.3a^{\star}\lesssim0.23\AM$. Multiplicity of ``critical''
wavevectors distributed over an extended region in ${\bf q}$-space is a
consequence of geometrical frustration in the pyrochlore lattice.

The suggested distribution in ${\bf q}$-space of strongly enhanced and slow
spin fluctuations is compatible with the angular averaged data derived from
inelastic neutron scattering on polycrystalline
samples.\cite{Krimmel99,Lee01,Murani04} Indeed, as seen from Fig.\
\ref{fig:xiq}, the next pronounced maximum of $\chi({\bf Q},0)$ is found
along the X-direction at $Q_{c2}^{X} \approx 1.4a^{\star}\approx
1\AM$. Together with the other equivalent directions, it makes up a 6-point
manifold but their partners at $Q_{c2}^{K,L}$ are strongly suppressed by
the form-factor. Therefore, this secondary manifold is not resolved in the
experiment.

Many metallic systems with enhanced spin fluctuations show dramatically
renormalized thermodynamic properties, for instance, a tendency to the
formation of heavy-fermion properties: a strongly enhanced linear term
$\gamma T$ of the specific heat and a concomitantly enhanced magnetic
susceptibility $\chi$. In many 4$f$- and 5$f$-derived heavy-fermion system
the spin fluctuations are soft (have low $\omega$) over large portion of
the reciprocal space, which provides a large value of $\gamma$. A different
case is the nearly antiferromagnetic metal Cr$_{0.95}$V$_{0.05}$ which is
close to incommensurate magnetic order. In contrast to the heavy-fermion
systems, Cr$_{0.95}$V$_{0.05}$ has a typical, unenhanced value of the
measured $\gamma$ which is explained\cite{Hayden00} to be due to the small
region in the momentum space occupied by the exchange enhanced soft spin
fluctuations in this material.

It is promising now to develop a spin fluctuation theory for LiV$_2$O$_4$
and estimate the contribution of spin fluctuations to the low-$T$ specific
heat. The theory is aimed at an approximate description of the low-$\omega$
behavior of the calculated $\chi({\bf q},\omega)$ in terms of overdamped
oscillators. One has to start with the following definition of the spin
fluctuation free energy\cite{Lonzarich86,Lonzarich89,Konno87}
\begin{equation}
F_{sf}=\frac{3}{\pi}\sum\limits_{\bf q} \int\limits_{0}^{\omega_c}
d{\omega}F_{osc}(\omega) \frac{z ({\bf q})\Gamma({\bf q})}
{\omega^{2}+\Gamma^{2}({\bf q})}\,,
\label{a19}
\end{equation}
where $F_{osc}(\omega)=k_{B}T\ln\left[1-\exp(-\hbar\omega/k_{B}T)\right]$
is the thermal part of the free energy of an oscillator; $z({\bf q})$ is
the spectral weight of a low-$\omega$ spin fluctuation mode and $\omega_c$
is the cutoff frequency. The specific heat is given by $C_{sf}(T)=
-T/N(\partial^2F_{sf}/\partial T^{2} )$, where $N$ is the number of V-atoms
in the system. In the limit $T\to 0$, simple calculation leads to the
familiar expression for the spin fluctuation contribution $\gamma_{sf}=
C_{sf}(T)/T$ to the specific heat coefficient $\gamma$:
\begin{equation}
\gamma_{sf}=\frac{k_{B}^2\pi}{N}\sum\limits_{\bf q} 
\frac{z({\bf q})}{\hbar\Gamma ({\bf q})},  
\label{a20}
\end{equation}
where the summation is over the cubic BZ.

Our aim now is to propose a reliable simple model describing variations of
$z ({\bf q})$ and $\Gamma ({\bf q})$ in ${\bf q}$-space. The model adopts
the most essential low-$\omega$ properties of $\chi ({\bf q},\omega )$
calculated for LiV$_2$O$_4$. We start by partitioning the ${\bf q}$-space
into three regions: (I) the interior of the surface depicted in Fig.\
\ref{fig:qsurf}, i.e., $|{\bf q}| <(|{\bf q}_c|-\delta q/2)$,
(II) the ``critical'' region $(|{\bf q}_c|-\delta q/2)<|{\bf q}|
<(|{\bf q}_c|+\delta q/2)$, and (III) the periphery, $|{\bf q}|
>(|{\bf q}_c|+\delta q/2)$, of the cubic BZ.  Within the layer (II)
we approximate $\Gamma ({\bf q} )$ and $z ({\bf q})$ by constants, $\Gamma$
and $z_{II}<1$, respectively. Then, in the region (I) $\Gamma_{I} ({\bf q}
)=(q/q_1)\Gamma$ and $z_{I} ({\bf q})=1-( 1-z_{II})q/q_1$, which provides a
piecewise continuity between regions (I) and (II). Here $q_1\simeq 0.5
a^{\star}$ and $q_2\simeq 0.8a^{\star}$ are minimal radii of the lower
(Fig.\ \ref{fig:qsurf}) and the upper bound surfaces for the ``critical''
region.  Finally, as suggested from the analysis of Im$\chi({\bf
q},\omega)$ at large ${\bf q}$, in (III) the weight $z({\bf q})$ decreases
fast to zero. Therefore, this contribution is neglected below. After
performing the integration in (\ref{a20}) we arrive at the following
expression
\begin{equation}
\gamma_{sf} = \frac{\pi^2}{32}k^2_B\left[\left(\frac{q_1}{a^{\star}}\right)^3 +
2z_{II}\left(\frac{q_2}{a^{\star}}\right)^3\right]\frac{1}{\hbar\Gamma}. 
\label{a21}
\end{equation}
Factor 3/2 involved in Eq.~(\ref{a21}) takes into account a deviation of
the ``critical'' surface shape from a spherical one.

An analysis of the calculated Im$\chi({\bf q},\omega)$ leads us to a rather
rough estimate $z_{II}\approx1/4$, while the reported
\cite{Krimmel99,Lee01,Murani04} low-$T$ values of $\hbar\Gamma$ in the
``critical'' region of ${\bf q}$ fall within the limits of
0.5meV$<\hbar\Gamma<$5meV. By substituting these parameters into
(\ref{a21}) and recalling that there are two V-atoms in the formula unit,
we obtain the estimate for $\gamma_{sf}$ as 300$>\gamma_{sf}>$30 in units
of mJ/K$^2$mol. This estimate shows that slow spin fluctuations over an
extended region in ${\bf q}$-space may explain the size of the enhanced
specific heat coefficient in LiV$_2$O$_4$.

It is instructive to estimate the Sommerfeld-Wilson ratio $R_{\text{W}} =
\pi^2 k_{\text{B}}^2 \chi_S (T=0)/(3 \mu_{\text{B}}^2 \gamma)$, where
$\mu_{\text{B}}$ is Bohr magneton.  Taking the upper bound for $\gamma$
coefficient as $\gamma\simeq$300 mJ/K$^2$mol and the value
$\chi_S(T=0)=\chi(0,0)$ calculated for ${\mathcal{K}}=0.95 {\mathcal{K}}_c$
in our theory, one obtains $R_{\text{W}} \approx$5. This result is in a
contrast with the estimate $R_{\text{W}}\sim$0.1 obtained within the RPA
theory\cite{Yamashita03} where both spin and orbital critical fluctuations
were assumed to contribute.

\section{Conclusions}
 
The enhanced dynamic spin susceptibility of the multi-band paramagnetic
spinel LiV$_2$O$_4$ was calculated on the basis of the actual LDA band
structure of this metallic system. It was shown that the complexity of the
band structure results in an unenhanced spin susceptibility that displays a
key information about spin fluctuations in this material, namely, the
position in ${\bf Q}$-space of dominant spin correlations.  The calculated
moduli of these ``critical'' wave-vectors ${\bf Q}_c$ located (at $T=0$)
obviously far away from the Brillouin zone center are in a good agreement
with experiment.

The susceptibility enhancement due to electron interaction is described and
calculated in the RPA approach adopted to the actual multi-band system with
nearly degenerate $d$-orbitals. The most substantial approximation we made
is the neglect of the orbital dependence and the off-diagonal matrix
elements of the matrix ${\mathcal{K}}$ defined by Eq.~(\ref{a14}). A
wave-vector dependence of ${\mathcal{K}}$ is also omitted, which, we
believe, is less crucial since the vanadium $d$-orbitals are well localized
and strong on-site electronic correlations dominate.  The approach
developed above may be extended in the following ways. Provided a
particular model of electron interaction is chosen and parameterized
suitably, matrix elements of the ${\mathcal{K}}$ matrix entering the
expression (\ref{a15}) may be evaluated.\cite{Bass96} The resulting
few-parameter theory then can be put on a quantitative ground by comparing
the calculated model results with available experimental data.
 
Despite the approximations used in the description of electron correlations
in a multi-band electronic system like LiV$_2$O$_4$, we believe that the
approach developed here enabled us to catch, in accord with experimental
observations, the most essential effect of correlations: enormous
magnification and slowing down of spin fluctuations at ``critical'' wave
vectors ${\bf Q}_{c}$.

To our opinion, the obtained value of the critical coupling constant
${\mathcal{K}}_c$=0.49 eV, at which a magnetic instability may occur in
LiV$_2$O$_4$, is rather low, which means that a generalized Stoner
criterion can be easily fulfilled. This finding is in agreement with the
discussion in Refs.\ [\onlinecite{Eyert99,Singh99}]. There, the authors
performed spin-polarized LDA band structure calculations and noted a close
proximity of the paramagnetic LiV$_2$O$_4$ to different types of magnetic
instabilities.

\begin{acknowledgments}
One of the authors, V.Yu., acknowledges support from Deutsche
Forschungsgemeinschaft under SFB 463.
\end{acknowledgments}


\end{document}